\documentclass[]{article}
\usepackage{textcomp}
\usepackage{amsmath,amssymb,amsfonts}
\usepackage{algorithmic}
\usepackage{graphicx}
\usepackage{textcomp}
\usepackage{hyperref}
\usepackage{float}
\usepackage{subfig}
\usepackage{breqn}
\usepackage{multirow,multicol}
\usepackage{makecell}
\usepackage{cellspace}
\usepackage{authblk}
\usepackage[a4paper, total={6.5in, 9in}]{geometry}

\def\BibTeX{{\rm B\kern-.05em{\sc i\kern-.025em b}\kern-.08em
    T\kern-.1667em\lower.7ex\hbox{E}\kern-.125emX}}

\newcommand{\beginsupplement}{%
        \setcounter{table}{0}
        \renewcommand{\thetable}{S\arabic{table}}%
        \setcounter{figure}{0}
        \renewcommand{\thefigure}{S\arabic{figure}}%
     }

\begin{document}
\title{TricycleGAN: Unsupervised Image Synthesis and Segmentation Based on Shape Priors}

\author[1]{Umaseh Sivanesan}
\author[1,2]{Luis H. Braga}
\author[1,3,4]{Ranil R. Sonnadara}
\author[1,3,4]{Kiret Dhindsa\thanks{Corresponding Author: kiretd@gmail.com}}

\affil[1]{Department of Surgery at McMaster University, Hamilton, Ontario, Canada}
\affil[2]{Division of Paediatric Urology, McMaster Children's Hospital, Hamilton, Ontario, Canada}
\affil[3]{Research and High Performance Computing, McMaster University, Hamilton, Ontario, Canada}
\affil[4]{Vector Instutite for Artificial Intelligence, Toronto, Ontario, Canada}

\maketitle

\begin{abstract}
Medical image segmentation is routinely performed to isolate regions of interest, such as organs and lesions. Currently, deep learning is the state of the art for automatic segmentation, but is usually limited by the need for supervised training with large datasets that have been manually segmented by trained clinicians. The goal of semi-superised and unsupervised image segmentation is to greatly reduce, or even eliminate, the need for training data and therefore to minimze the burden on clinicians when training segmentation models. To this end we introduce a novel network architecture for capable of unsupervised and semi-supervised image segmentation called TricycleGAN. This approach uses three generative models to learn translations between medical images and segmentation maps using edge maps as an intermediate step. Distinct from other approaches based on generative networks, TricycleGAN relies on shape priors rather than colour and texture priors. As such, it is particularly well-suited for several domains of medical imaging, such as ultrasound imaging, where commonly used visual cues may be absent. We present experiments with TricycleGAN on a clinical dataset of kidney ultrasound images and the benchmark ISIC 2018 skin lesion dataset. 

\date{}
\end{abstract}

\section{Introduction}
\label{sec:introduction}
In vivo medical imaging is an important and commonly used technique in a wide variety of clinical applications. However, analysis is often made challenging due to the poor quality of images obtained with current imaging technologies. In order to take advantage of these technologies, current medical practice relies heavily on clinicians who are highly specialized in interpreting medical images \cite{Ravi2019,Tajbakhsh2019}. 

Segmentation, or identifying and isolating regions of interest (ROIs) from an image, is often a key step in medical image analysis. For both human and machine readers, segmentation aids in the extraction of clinically important features by identifying relevant tissues and/or objects, such as an organ or a lesion, without interference from non-relevant tissues captured in the image \cite{Guo2018}. Since manual segmentation, i.e., having appropriately trained clinicians segment images by hand, is time-consuming, expensive, and subjective, significant effort has been put into developing algorithms that can automatically provide accurate and reliable segmentations for downstream analysis \cite{Chowdhary2020}. This has lead to the growth of the subfield of semantic segmentation, which is image segmentation via the association of pixels with classification labels (e.g., kidney vs. not kidney).

Here we present a novel approach to image segmentation that can be trained unsupervised or semi-supervised, thus greatly reducing, or even eliminating, the need for manually segmented images. Inspired by the way trained clinicians use anatomical knowledge of the shapes of ROIs and their internal structures when performing manual segmentation, we rely on shape priors to generate synthetic labelled data that can be used to train a segmentation model. Shape priors are built in by generating synthetic medical images from ROI shape templates, making it possible to learn mappings between the generated segmentation-medical image pairs. We expand upon the rationale and methodology for this approach in Section \ref{sec:modelling}. 

Our new architecture, TricycleGAN, is comprised of three Generative Adversarial Networks (GANs), specifically, three CycleGANs \cite{Zhu2017} that have been chained together. Each CycleGAN learns one of three mappings: a mapping between original medical images and edge maps derived from those medical images, a mapping between edge maps and segmentation maps for the regions of interest contained in those medical images, and a mapping between segmentation maps and medical images. As a whole, TricycleGAN unifies these three subnetworks using an adapted cycle consistency loss and thus learns to segment medical images by learning to generate realistic image-edge-segmentation triplets. As is explained further in Section \ref{sec:modelling}, the inclusion of edge maps into the process provides an intermediate step that introduces sources of variance and complexity that allows TricycleGAN to translate simplistic ROI templates into realistic medical images while retaining the necessary shape information needed to predict a segmentation map. 

To demonstrate the value of this novel approach, we evaluate our method using two distinct datasets: a dataset of ultrasound images for which the task is to segment the kidney, and the ISIC 2018 Skin Lesion Analysis competition dataset, for which the task is to segment skin lesions in dermoscopic images.

\section{Related Work}

Prior to deep learning approaches to automated image segmentation, techniques based on edge detection \cite{Canny1987}, region growing \cite{Adams1994}, contour modelling \cite{Kass1988}, and texture analysis \cite{Manjunath1991} relied on detecting statistical difference among objects in an image using explicitly defined constraints and criteria. These methods can often perform pooly in medical imaging applications because typically used metrics, such as contrast gradients, are not robust enough to overcome the challenges imposed by noisy images, e.g., in ultrasound imaging, and artifacts, e.g., surrounding tissues. 

A current strategy for image segmentation aims to overcome such limitations using supervised deep neural networks that can learn to identify distinct objects and ROIs in images by discovering combinations of discriminative features given exposure to numerous labelled samples. Convolutional neural networks (CNNs) have been particularly successful for semantic segmentation and a wide variety of other computer vision applications \cite{Voulodimos2018, Yoo2015, Garcia2017, Lu2017}. Notably, Fully Convolutional Networks (FCNs)  omit the fully connected layers used to group pixels by class label in standard CNNs in favour of deconvolution layers used to estimate segmentation probability maps for entire images \cite{Long2015}. Encoder-decoder networks, first introduced by deconvolving a pretrained VGG16 CNN architecture \cite{Simonyan2014}, are based on a similar principle. CNNs have also been paired with Conditional Random Fields (CRFs) to better take advantage of the spatial correlations among pixels belonging to the same object \cite{Lin2016}. 

Variations on these segmentation-oriented CNNs have also played an important role in clinical applications of medical imaging \cite{Moeskops2016, Zhang2015, Kleesiek2016, Dou2016, Brosch2016, Cirecsan2013}. The U-net \cite{Ronne2015}, a type of FCN, has performed well in a variety of medical image segmentation tasks (e.g., \cite{Christ2016}), including for 3D segmentation tasks \cite{Cciccek2016}. Various extensions have been developed, including to incorporate an attention mechanism \cite{Oktay2018}, or by altering the initialization of the model \cite{Iglovikov2018}. 

While successful in many areas, CNN-based approaches can be prone to poor performance under certain conditions that are often observed in medical imaging applications. One issue is that obtaining spatially and semantically contiguous solutions usually requires significant post-processing of the resulting segmentation maps. In the presence of artifacts and distractors, such as the presence of multiple salient organs when only one is the desired ROI, such methods can result in large errors. Fortunately, recent work has begun to address this issue \cite{Havaei2017, Kamnitsas2017, Pereira2016}. However, another significant limitation for real-world use is that deep learning approaches in general usually require large amounts of training data, which usually means hundreds or thousands of ground-truth segmentations that must be hand-drawn by radiologists outside of their clinical duties. Although some success has been seen with unsupervised CNN approaches under some conditions, e.g., with W-net \cite{Xia2017} (typically only successful for segmentation tasks with distinct non-overlapping objects) or DeepCo3 \cite{Hsu2019} (relying on feature similarity among multiple instances of same-class objects in an image), unsupervised solutions for some medical applications may require an altogether different approach.

The use of GANs has been proposed as one way to address the need for ground truth segmentation labels. GANs learn to perform image-to-image translation by training with pairs of images \cite{Isola2017}, and therefore can be appropriated for segmentation tasks by training them with image-segmentation map pairs in supervised \cite{Luc2016,Xue2018,Son2017,Yang2017,Moeskops2016} or semi-supervised \cite{Hong2015} fashions. A useful function of GANs is their ability to augment training datasets by generating realistic synthetic data, thus requiring relatively smaller amounts of labelled training data \cite{Shin2018, Souly2017, Guibas2018}. To adapt GANs to learn segmentation in an unsupervised manner, recomposition approaches have been used. For example, SEIGAN can be used to segment foreground objects by exploiting distinct feature statistics between background and foreground objects and similar feature statistics between similar backgrounds \cite{Ostyakov2019}. Another recent approach, ReDO, uses scene decomposition to separate objects in an image based on the assumption that each object is distinct with respect to specific properties, such as colour and texture \cite{Chen2019}. 

Unsupervised approaches for image segmentation make reasonable assumptions about ways in which the desired ROI is distinct from other objects in an image. However, typically relied upon features, such as colour, texture, or brightness, may not carry over well to many medical imaging domains. In the case of organ segmentation, as illustrated with the kidney ultrasound dataset used in this study, a clinician must rely primarily on a priori anatomical knowledge and experience in order to estimate the contours of the kidney when a clear boundary is not visible. This task is made especially difficult by the fact that the kidney may not even be the most salient object in the image, leading even clinicians from other domains to find the task very challenging. 

To fill the gap left by previous methods, TricycleGAN is aimed at overcoming these challenges by incorporating a shape prior for the ROI to generate synthetic image-segmentation map pairs using only unlabelled medical images, simulating the use of a priori anatomical knowledge that a radiologist would use to perform similar segmentation tasks. TricycleGAN is an extention of CycleGAN \cite{Zhu2017} that uses three generators: the first translates between segmentation maps and edge maps, the second translates between edge maps and medical images, and the third uses the cycle consistence loss to translate medical images back into segmentation maps. An illustration of the TricycleGAN pipeline is given in Figure \ref{fig:pipeline}, and examples showing the output at every major step of the pipeline using the ISIC 2018 skin lesion dataset are shown in Figure \ref{fig:derm_steps}.


\section{Data and Preparation}
\subsection{Dataset 1: Renal Ultrasound Images}
\label{sec:kidney_data}
We use a dataset of renal ultrasound images developed for prenatal hydronephrosis, a congenital kidney disorder marked by excessive and potentially dangerous fluid retention in the kidneys. The dataset consists of 2492 2D sagittal kidney ultrasound images from 773 patients across multiple hospital visits.The evaluation set consists of 438 images that have been manually segmented by a trained surgical urologist. After removing training images taken from the same patients represented in the evaluation set, 918 unlabelled images remain for training. During training, random samples of 20\% of the images were used for validation. Each grade of hydronephrosis is represented in the evaluation set approximately evenly. 

This is a difficult dataset for image segmentation due to poor image quality, unclear contours of the kidneys, and the large variation introduced by different degrees of the kidney disorder called hydronephrosis (see Supplementary Figure \ref{fig:kidney_grades}). In addition, a major challenge of this dataset is that the two most salient boundaries are the outer ultrasound cone inherent to ultrasound imaging with a probe, and the dark inner region of the kidney, which is caused by fluid retention in hydronephrosis. As neither of these are the desired ROI, both are misleading with respect to segmenting the kidney. Further details concerning this dataset can be found in \cite{Dhindsa2018,Sivanesan2019,Smail2020}.

\subsection{Dataset 2: Skin Lesion Segmentation}
\label{sec:derm_data}
We use the ISIC 2018 Lesion Boundary Segmentation Challenge dataset \cite{Codella2019,Tschandl2018} to more directly compare TricycleGAN with other approaches. The provided training set of 2075 images, 100 validation images, and 519 evaluation images are used for the intended purposes in this study. By showing that TricycleGAN is also successful on this benchmark dataset, we show that it is not limited to only one domain and imaging modality. In addition, we show one method by which TricycleGAN can be adapted to also take advantage of commonly used features, such as colour.

\subsection{Image Preprocessing}
We follow a similar methodology used for preprocessing renal ultrasound images described in \cite{Dhindsa2018}. We crop the images to remove white borders, despeckle them to remove speckle noise caused by interference with the ultrasound probe during imaging \cite{Tay2010}, and re-scale to 256$\times$256 pixels for consistency. We remove text annotations made by clinicians using the pre-trained Efficient and Accurate Scene Text Detector (EAST) \cite{Zhou2017}. We then normalize the pixel intensity of each image to be from 0 to 1 after trimming the pixel intensity from the 2\textsuperscript{nd} percentile to the 98\textsuperscript{th} percentile of the original pixel intensity across the image. In addition, we enhance the contrast of each image using Contrast Limited Adaptive Histogram Equalization with a clip limit of 0.03 \cite{Pizer1987}. Finally, we normalize the images by the mean and standard deviation of the training set during cross-validation. 

We perform no preprocessing for the ISIC skin lesion images other than to resize them to 256 $\times$ 256 pixels.

\section{TricycleGAN}
\label{sec:modelling}
\subsection{Rationale}
Analogously to how other image features are exploited for semantic segmentation, we assume that the shape of an ROI for a particular image segmentation task is statistically more similar for within-class objects than it is for background objects. In order to make use of this assumption, we additionally assume that the boundary of the ROI is sufficiently prominent in the image to appear in its corresponding edge map (though as mentioned previously in Section \ref{sec:kidney_data}, this does not require that the ROI boundary is the most prominent boundary in the image, and does not preclude partial occlusion). Based on these two assumptions, we make use of a basic template shape that is appropriate for representing the ROIs of a given class of objects. For example, both kidneys and skin lesion ROIs are roughly elliptical. Since TricycleGAN adds the required complexity to make the resulting synthetic images realistic on its own, constructing a template can be as simple as drawing an ellipse with randomized shape, size, and location. Details of this part of the process are given in Section \ref{sec:synth_seg}. 

Note that by constructing simple ROIs explicitly and generating synthetic images from them, we also provide our model with a ground truth segmentation label for the synthetic images. It is for this reason that we do not generate the underlying ground truth with a generative model. Instead, the level of variance and complexity required to generate realistic images in the application domain are introduced by the downstream generators in TricycleGAN.

In addition to the potentially large variation in ROI morphologies within a class, a common challenge with medical imaging segmentation is the presence of artifacts and distractors (i.e., objects or anatomical features that also appear prominently in edge maps, but which are not part of the intended ROI). We overcome both of these challenges simultaneously by using a generative model to learn the statistics of edge maps extracted from real images, which can then be used to construct realistic edge maps from generated template ROIs (see Section \ref{sec:seg2edge}). By then constructing synthetic images in the application domain from the synthetic edge maps, TricycleGAN generates segmentation map - edge map - medical image triplets that can be used for segmentation training. 

The reason for using edge maps as an intermediary step is that edge maps can be easily extracted from real training data using known statistical methods, providing a training set of real image - edge map pairs that can be used to learn a mapping with supervision. Since segmentation maps cannot be extracted from the real images, a direct mapping cannot be learned. Instead, as described in Section \ref{sec:seg2edge}, we introduce a patch-occlusion method for learning the statistics of real edge maps and use this to generate realistic edge maps from synthetic segmentation maps, thus completing the loop from synthetic segmentation map to synthetic edge map to synthetic image and back to a predicted segmentation map with the required complexity. 

\begin{figure}
	\centering
	\includegraphics[width=\textwidth]{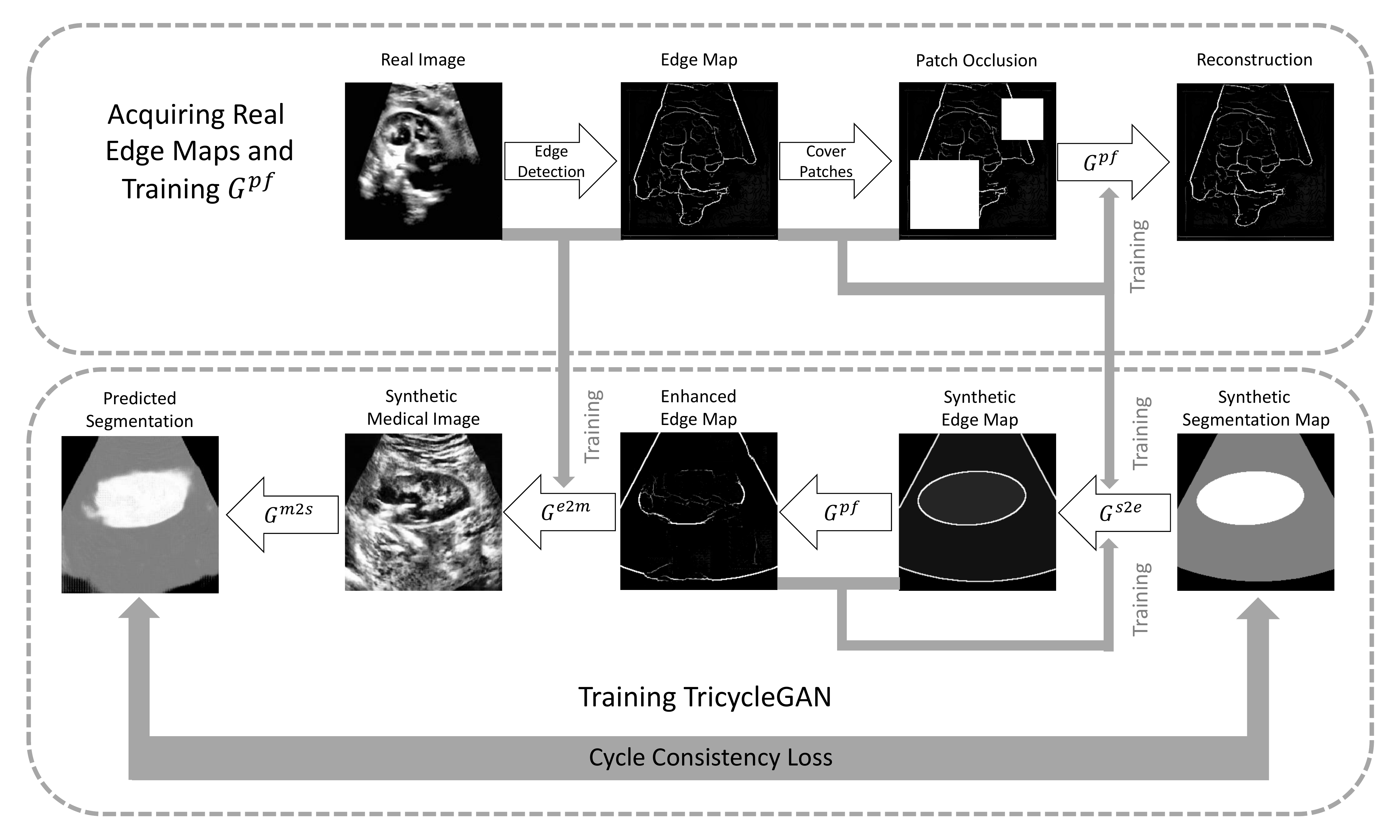}
	\caption{The TricycleGAN training pipeline.}
	\label{fig:pipeline}
\end{figure}

\begin{figure}
	\centering
	\includegraphics[width=0.45\textwidth]{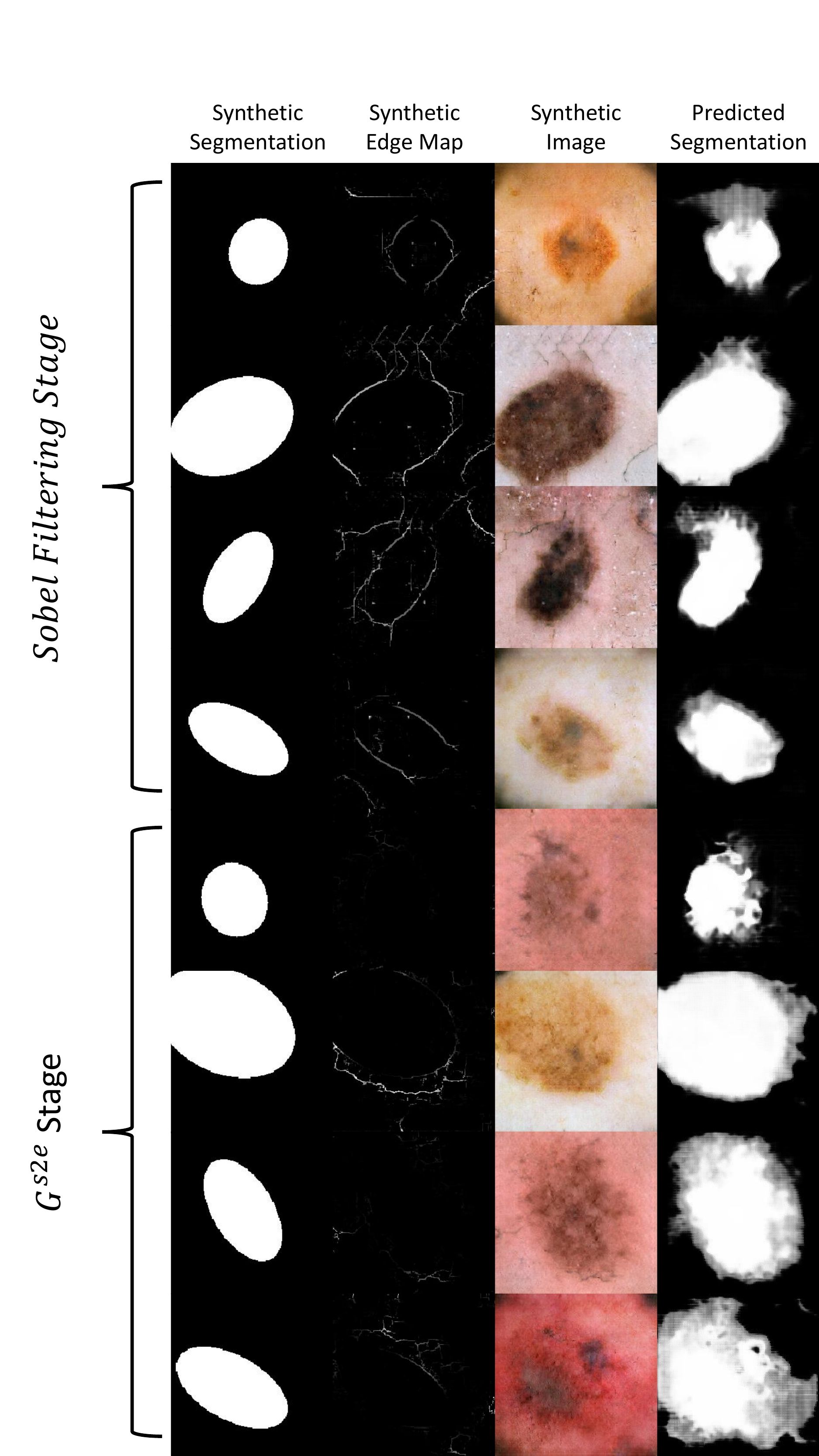}
	\caption{Examples showing the output of each major step of the program, including synthetic segmentation map construction and the output of each generator.}
	\label{fig:derm_steps}
\end{figure}

\subsection{Constructing Synthetic Segmentation Maps}
\label{sec:synth_seg}
The primary concern when construction ROI templates, or synthetic segmentation maps, to be used as ground truth segmentation maps is to capture the distribution of shapes and locations that can be expected in the real data. For many applications this is relatively simple to do.

To generate an ROI, we create a randomized ellipse with a random origin (horizontally and vertically offset from the center by up to 1/8 of the image size), rotation (any angle), and major and minor axes (major axis of length between 1/4 and 1/2 of the image length, and minor axis between 0.5 and 0.9 times the length of the major axis). The ellipse itself serves as the ground truth segmentation mask.

For the ultrasound dataset we require the addition of the ultrasound cone, as its outline is even more prominent than the desired kidney ROI. The cone is also randomly generated by creating a randomized bottom curve (a partially complete ellipse) and a triangle extending beyond the upper boundary of the image (so that it is cut off at the top). All pixels of the generated kidney ROI not lying within the generated ultrasound cone is removed to simulate the common occurrence of only partially captured kidneys. Similar adjustments in constructing ROI templates can be made for other applications where prominant shapes may serve as distractors.

\subsection{Extracting Edge Maps from Real Images}
\label{sec:edge_real}
Edge maps that are extracted from real images are used to train the synthetic edge map generator. We extract edge maps from real medical images using the VGG16 model \cite{Simonyan2014} with Richer Convolutional Features pretrained as in Liu et al. \cite{Liu2017}. As recommended for this approach, we further fine-tune the output edge maps using non-maximum suppression with Structured Forests for edge thinning \cite{Dollar2014}. 

\subsection{Generator 1: Segmentation Map to Edge Map Translation}
\label{sec:seg2edge}
To generate images with structural variation resembling that observed in the real data, we first convert the synthetic segmentation maps into realistic edge maps, complete with artifacts and natural irregularities, using the first generative model in TricycleGAN, $G^{s2e}$. As with all of the generators used in TricycleGAN, $G^{s2e}$ is based on the \emph{pix2pix} architecture defined in \cite{Isola2017}. 

To train generator $G^{s2e}$ to produce the appropriate variation seen in the real data, we frame this problem as an image completion problem (e.g., \cite{Liu2017b}) and make use of our previously extracted real edge maps. However, since $G^{s2e}$ takes as input the very simplistic synthetic segmentation maps without any corresponding realistic edge maps (since the edge maps extracted from real images do not have paired segmentation maps), on its own it learns to produce very simple edge maps. This is partially because the use of the cycle consistency loss used to train TricycleGAN (see Section \ref{sec:loss} below) encourages less complex edge maps. We therefore incorporate an additional generator based on the \emph{pix2pix} architecture, denoted $G^{pf}$ (a patch-filling generator), which enhances the complexity of the edge maps produced by $G^{s2e}$ by adding artifacts and a greater degree of irregularity. Therefore, edge maps produced by $G^{s2e}$ are patch-occluded and then passed into $G^{pf}$ for added complexity. 

We pretrain $G^{pf}$ on the \emph{edges2handbags} dataset \cite{Zhu2016} before further training it on the edge maps extracted from the training set of real images after partial occlusion using randomly generated square masks. For each extracted edge map, up to 10 non-overlapping square masks with length between 1/8 and 1/2 of the image length are used. Once this is done, the weights of $G^{pf}$ are frozen allowing it to produce the artifacts and necessary complexity to emulate realistic edge maps, which $G^{s2e}$ does not encounter, without loss of complexity due to the cycle consistency loss. Additionally, this allows us to use $G^{pf}$ to help train $G^{s2e}$ by serving as a regularizer. Specifically, the $L_1$-loss used to regularize $G^{s2e}$ is computed by taking the difference between the output generated by $G^{s2e}$ and $G^{pf}$, thereby encouraging $G^{s2e}$ to increase the complexity of its output to provide $G^{pf}$ a better starting point for enhancement, and countering some of the simplicity encouraged by the cycle consistency loss. 

A natural question is why TricycleGAN cannot simply use $G^{pf}$ without $G^{s2e}$. In practice, $G^{pf}$ requires its weights to be frozen in order to maintain its ability to generate artifacts. Therefore, it does not learn to contribute to minimizing the overall loss of the network, thus breaking unification of the overall model and ultimately causing TricycleGAN to perform suboptimally. If the weights of $G^{pf}$ are allowed to change, the cycle consistency loss causes it to gradually simplify its edge maps until the desired complexity is no longer produced. Alternatively, using $G^{s2e}$ without influence from $G^{pf}$ results in overly simplistic edge maps that fail to provide enough challenge to the later generators of TricycleGAN. The required balance of complexity and optimal training for TricycleGAN is achieved by using $G^{pf}$ both as an augmenter of the output of $G^{s2e}$ and as a regularizer for its training.


In practice, while $G^{s2e}$ undergoes training with $G^{pf}$, we begin by using simple Sobel filtering to convert segmentation maps into edge maps so that meaningful outputs can be generated by downstream generators. After 250 epochs of training TricycleGAN, the probability of using the output of $G^{s2e}$ versus Sobel filtering is linearly ramped down until it reaches zero at epoch 500. From there, Sobel filtering is discontinued and $G^{pf}$ takes only the output of $G^{s2e}$ as input. In this way, the downstream generators of TricycleGAN can still learn to produce meaningful outputs while $G^{s2e}$ learns to initially be able to produce edge maps, and TricycleGAN as a whole can continue to improve as $G^{s2e}$ learns to produce edge maps with greater complexity than provided by Sobel filtering.

Finally, $G^{s2e}$ is ready to be used to construct realistic edge maps from the synthetic segmentation maps with artifacts and added complexity contributed by $G^{pf}$. Figure \ref{fig:derm_steps} provides examples of generator outputs at each stage of the pipeline comparing $G^{pf}$ ouputs during the early stages of training when Sobel filtering is used to later stages of training when $G^{s2e}$ is used.

\subsection{Generator 2: Edge Map to Medical Image Translation}
\label{sec:image_synth}
The second generator in TricycleGAN, $G^{e2m}$ translates edge maps into realistic images. As before, we use a \emph{pix2pix} model here. $G^{e2m}$ is trained using the extracted edge map and real image pairs obtained via our training data while also training through the cycle consistency loss as part of TricycleGAN when synthetic images are used as input. Once trained, the generator is used to translate the edge maps previously constructed from $G^{s2e}$ into realistic synthetic medical images.

\subsection{Generator 3: Medical Image to Segmentation Map Translation}
\label{sec:image_seg}
The final generator in TricycleGAN, $G^{m2s}$ performs image segmentation by translating a real or synthetic image into its corresponding segmentation map. Again, we use the \emph{pix2pix} architecture here. The model is trained using the synthetic segmentation maps and their corresponding synthetic medical images created by $G^{e2m}$. Since no real image - segmentation map pairs are available for the training data, $G^{m2s}$ is only trained using synthetic images. Thus, it undergoes considerably less training than the other generators. The specific loss functions used to train each generator are given below.

\subsection{Training Scheme and Loss Functions}
\label{sec:loss}
\subsubsection{Training Pipeline}
TricycleGAN is trained using a combination of unlabelled real images and labelled synthetic images. Since we extract edge maps from the real images, image - edge map pairs are available. Using the patch occlusion method, the real edge maps are used to pretrain $G^{pf}$, as described in Section \ref{sec:seg2edge}. When training TricycleGAN with the weights of $G^{pf}$ frozen, the output of $G^{pf}$ is used to compute the $L_1$ loss for $G^{s2e}$. The real image - edge map pairs are also used to train $G^{e2m}$. However, with no segmentation map available, no training signal is computable for $G^{m2s}$ when TricycleGAN trains with real images. Instead, every 20th image is generated using a synthetic segmentaiton map that is passed through the full network, meaning that $G^{m2s}$ trains at a slower rate compared to the other generators.

To ensure that generator $G^{m2s}$ learns to generate accurate segmentation maps, we constrain the output space of the network by incorporating the cycle consistency loss first introduced into adversarial learning with CycleGAN. This pushes $G^{m2s}$ to translate an image back into the original image input into the network during training, i.e., the synthetic segmentation map \cite{Zhu2017}. A second motivating factor for using the cycle consistency loss is that it does not assume that the embedding spaces of the input and output are the same (in our case, segmentation maps are almost guaranteed to lie in a lower-dimensional embedding space than the medical images themselves), which provides some desired flexibility in TricycleGAN. 

When training using synthetic images, the cycle consistency losses for each trainable generator is computed after the initial segmentation map is passed through the entire network, which provides a training signal for $G^{m2s}$. The predicted segmentation map generated by $G^{m2s}$ is then looped back to the start of the network and passed through each generator again, and the difference between their outputs on the second pass versus their outputs on their first pass are used to compute the cycle consistency loss (with the caveat that $G^{s2e}$ uses the output of $G^{pf}$). This allows TricycleGAN to take as much advantage of the real data as possible, while using synthetic data to train the chain of generators to function as a cohesive network that serves a specific purpose. Since TricycleGAN uses three trainable generators instead of CycleGAN's two, we use an adapted cycle consistency loss that is split into three parts; one for each generator. The training pipeline is illustrated in Figure \ref{fig:pipeline}, and further details of how each generator is trained as part of the network and how the adapted cycle consistency loss is defined are provided below.

\subsubsection{Training Generator $G^{s2e}$}
To train generator $G^{s2e}$, we start with the adversarial loss defined in \cite{Isola2017}:

\begin{dmath}
\min_{\theta_G}\max_{\theta_D}\mathcal{L}_{GAN}(\theta_G,\theta_D) + \gamma\mathcal{L}_{L_1}(\theta_G).
\end{dmath}

Here, $\theta_G$ and $\theta_D$ are parameters for a generator $G$ and a discriminator $D$ paired in a GAN.  

\begin{dmath}
    \min_{\theta_G}\max_{\theta_D}\mathcal{L}_{GAN}(\theta_G,\theta_D) = \mathbb{E}_{x\sim P_X}\left[log(D(x))\right] + \mathbb{E}_{z\sim P_Z}\left[log(1-D(G(z)))\right],
\label{eq:ganloss}
\end{dmath}

is the conventional adversarial loss \cite{Goodfellow2014}, where $x$ is a real image from the training set $X$ with unknown distribution $P_X$, and $z$ is a random noise vector from a Gaussian distribution $P_Z$, and 

\begin{dmath}
\mathcal{L}_{L_1}(\theta_G) = \mathbb{E}_{x\sim P_X, z\sim P_Z}\lVert x-G(z)\rVert_1
\label{eq:l1loss}
\end{dmath}

is simply the $L_1$ loss between the generated image and the target image. When training with real images, the target image for $G^{s2e}$ is defined as the real extracted edge map, and the generated image is $G^{pf}(G^{s2e}(x))$, as described in Section \ref{sec:seg2edge}. Full details of the construction of this loss function and the \emph{pix2pix} network architecture can be found at \cite{Isola2017}.

To train the generators in TricycleGAN to work together as a cohesive network, we also include a cycle consistency loss for each generator. For an edge map $x_1$,  cycle consistency is satisfied for $G^{s2e}$ if $x_1 \rightarrow G^{pf}(G^{s2e}(x_1)) \rightarrow G^{e2m}(G^{pf}(G^{s2e}(x_1))) \rightarrow G^{m2s}(G^{e2m}(G^{pf}(G^{s2e}(x_1)))) \rightarrow G^{pf}(G^{s2e}(G^{m2s}(G^{e2m}(G^{pf}(G^{s2e}(x_1)))))) \approx x_1$. The cycle consistency loss portion for $G^{s2e}$ is defined as 


\begin{dmath}
    \mathcal{L}_{Cycle}^{s2e}(G^{s2e},G^{e2m},G^{m2s}) = \mathbb{E}_{x_1\sim P_{X_1}}\lVert G^{pf}(G^{s2e}(G^{m2s}(G^{e2m}(x_1)))) - x_1\rVert_1.\
	\label{eq:cycle1}
\end{dmath}


${L}_{Cycle}^{s2e}$ is computed by passing the generated edge map through the generators in TricycleGAN to produce an image with $G^{e2m}$, leading to a predicted segmentation map generated by $G^{m2s}$, which is then used as input to produce another edge map with $G^{s2e}$. The original real edge map and the newly generated edge map produced after a cycle through TricycleGAN are used to compute the cycle consistency loss for $G^{s2e}$. 

\subsubsection{Training Generator $G^{e2m}$}
Generator $G^{e2m}$ is trained similarly to $G^{e2m}$. The adversarial loss and the $L_1$ loss are defined as before, except the target image for $G^{e2m}$ is the medical image domain. As $G^{e2m}$ starts at a different point in TricycleGAN's cycle, it contributes the second part of the total cycle consistency loss:

\begin{dmath}
    \mathcal{L}_{Cycle}^{e2m}(G^{s2e},G^{e2m},G^{m2s}) = \mathbb{E}_{x\sim P_X}\lVert G^{e2m}(G^{pf}(G^{s2e}(G^{m2s}(x_2)))) - x_2\rVert_1,
	\label{eq:cycle2}
\end{dmath}

where $x_2$ is a real image from the application domain and $y$ is the attempted reproduction of that image generated by $G^{e2m}$.

\subsubsection{Training Generator $G^{m2s}$}
 $G^{m2s}$ can only undergo training when a ground truth segmentation map is available, and thus only receives a training signal when a synthetic segmentation map is passed through TricycleGAN. Importantly, $G^{m2s}$ is not part of a GAN, and therefore there is no adversarial loss. Instead, it acts as a segmentation model, and thus is trained using its portion of the cycle consistency loss, the binary cross-entropy loss, and the Tversky loss. 
 
For $G^{m2s}$, we can define the cycle consistency loss between a ground truth segmentation $x_3$ and its predicted segmentation as 
 
\begin{dmath}
    \mathcal{L}_{Cycle}^{m2s}(G^{s2e},G^{e2m},G^{m2s}) = \mathbb{E}_{x\sim P_X}\lVert G^{m2s}(G^{e2m}(G^{pf}(G^{s2e}(x_3)))) - x_3\rVert_1.
	\label{eq:cycle3}
\end{dmath}
 
The binary cross-entropy is the sum of the cross-entropy loss for all $i$ pixels is defined as
\begin{dmath}
\mathcal{L}_{BCE} = -\sum_{i}(y_i\log(\hat{y}_i) + (1-y_i)\log(1-\hat{y}_i))
\label{eq:bce}
\end{dmath}
for pixel-wise ground truth labels $y_i$ and model predictions $\hat{y}_i$.

%

Finally, we add the Tversky loss, which is a generlization of the dice loss \cite{Abraham2019}, to promote segmentation accuracy. Given the Tversky similarity index for a two-class problem



\begin{dmath}
TI = \frac{\sum P_g(i)P_y(i) + \epsilon}{\sum P_g(i)P_y(i) + \alpha\sum Q_g(i) + \beta\sum Q_y(i) + \epsilon},
\label{eq:tversky}
\end{dmath}

where 

	\begin{equation}
	P_g{i} = \begin{cases}
 				1, & \text{if pixel $i$ is in the ground truth ROI} \\
				0, & \text{otherwise},
			\end{cases}
	\end{equation} 

$Q_g(i) = (1-P_g(i))P_y(i)$, and $Q_y(i) = P_g(i)(1-P_y(i))$.


The Tversky loss is simply $\mathcal{L}_{T} = 1-TI$. The parameters $\alpha$ and $\beta$ allow for shifting emphasis towards minimizing false positives or false negatives, depending on the class imbalance exhibited in the data. Here we use $\alpha=\beta=0.5$ to maintain a balanced emphasis.

\subsubsection{Total Loss for Training TricycleGAN}
To train generator $G^{s2e}$, we add together the cycle consistency losses for $G^{s2e}$ (Eq. \ref{eq:cycle1}), and $G^{e2m}$ (Eq. \ref{eq:cycle2}), along with its adversarial loss (Eq. \ref{eq:ganloss}) and its $L_1$ loss (Eq. \ref{eq:l1loss}) to obtain

\begin{dmath}
\mathcal{L}^{s2e} = \mathcal{L}_{GAN}(\theta_{G^{s2e}},\theta_{D^{s2e}}) + \lambda_1\mathcal{L}_{L_1}(\theta_{G^{s2e}}) + \mathcal{L}_{Cycle}^{s2e}(G^{s2e},G^{e2m},G^{m2s}) + \mathcal{L}_{Cycle}^{e2m}(G^{s2e},G^{e2m},G^{m2s}).\
\label{eq:s2e_loss}
\end{dmath}

Note that the sum of $\mathcal{L}^{s2e}_{Cycle}$ and $\mathcal{L}^{e2m}_{Cycle}$ is analagous to the sum of the forward and backward cycle consistency in CycleGAN. 

Similarly for $G^{e2m}$, the total loss is 

\begin{dmath}
\mathcal{L}^{e2m} = \mathcal{L}_{GAN}(\theta_{G^{e2m}},\theta_{D^{e2m}}) + \lambda_1\mathcal{L}_{L_1}(\theta_{G^{e2m}}) + \mathcal{L}_{Cycle}^{s2e}(G^{s2e},G^{e2m},G^{m2s}) + \mathcal{L}_{Cycle}^{e2m}(G^{s2e},G^{e2m},G^{m2s}).\
\label{eq:e2m_loss}
\end{dmath}

To train $G^{m2s}$, we combine the the cycle consistency losses for each trainable generator (Eq. \ref{eq:cycle1}, Eq. \ref{eq:cycle2}, and Eq. \ref{eq:cycle3}), the binary cross-entropy loss (Eq. \ref{eq:bce}) and the Tversky loss (Eq. \ref{eq:tversky}) to obtain

\begin{dmath}
\mathcal{L}^{m2s} = \mathcal{L}_{Cycle}^{e2m}(G^{s2e},G^{e2m},G^{m2s}) + \mathcal{L}_{Cycle}^{e2m}(G^{s2e},G^{e2m},G^{m2s}) + \mathcal{L}_{Cycle}^{m2s}(G^{s2e},G^{e2m},G^{m2s}) + \mathcal{L}_{BCE} + \mathcal{L}_{T}.\
\label{eq:m2s_loss}
\end{dmath}




For each of the above combined losses, $\lambda_1=100$ as recommended in \cite{Isola2017} and $\lambda_2=10$. 

\subsubsection{Training Parameters}
All generators were trained using the Adam optimizer with a learning rate of $2^{-4}$ and a batch size of $32$. Training was considered complete when $\mathcal{L}_{m2s}$ did not improve for 20 consecutive epochs. 

\subsubsection{Data Augmentation}
During training, images are altered slightly to introduce additional variance, thereby allowing the resulting models to learn additional robustness and improve performance. The alterations for both datasets are as follows: a random translation of up to 30 pixels along both axes, and a horizontal flip with a probability of 0.5. The following additional alterations are applied to the ISIC 2018 training data: a random rotation of $r\in\{0, \pi/2, \pi, 3\pi/2\}$, up to a 20\% change in brightness, up to a 50\% change in contrast, up to a 5\% change in hue, and up to a 50\% change in saturation. 

\begin{figure}[h]
	\centering
	\includegraphics[width=1.0\textwidth]{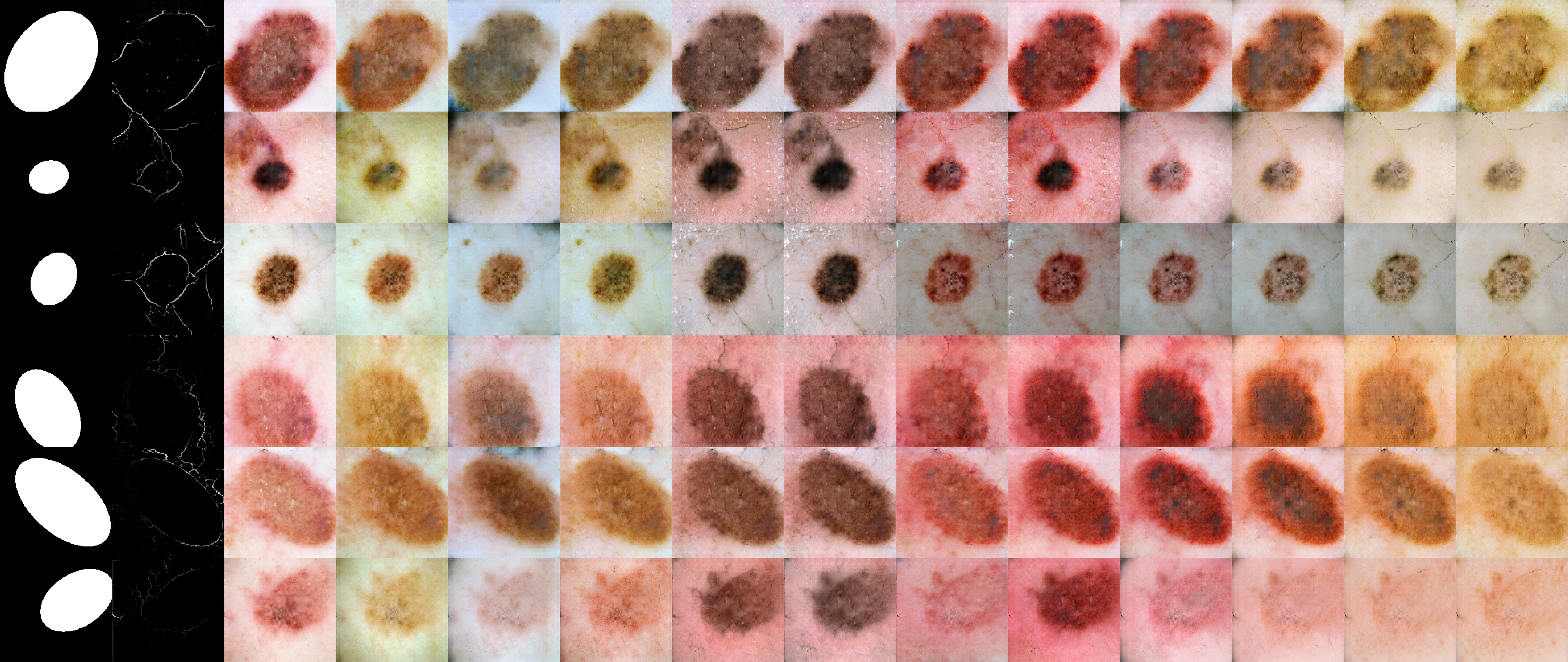}
	\caption{Examples of generated skin lesions. From left to right: synthetic segmentations, generated edge maps, followed by 12 style variations generated by applying 12 different styles with TricycleGAN generator $G^{e2m}$.}
	\label{fig:derm_gen_styles}
\end{figure}

We also use style transfer to provide additional data augmentation for the ISIC 2018 dataset. This allows the generators to produce a greater variety of skin lesions, particularly to better capture the statistics of different skin colours and artifacts (see Figure \ref{fig:derm_gen_styles}). This allows TricycleGAN to take advantage of colour features in addition to the usual shape features upon which it relies. Rather than treating each style variation as separate training images, styles were incorporated as an additional input channel (or dimension), enabling the networks to more effectively learn to produce the same segmentation for the various style variants of an image.

\section{Evaluation}
We evaluate our model using five standard metrics computed using the withheld evaluation images: the F1 score (also called the dice coefficient), specificity, sensitibity, intersection over union (IoU; also called the Jaccard Index), and pixel-wise classification accuracy (pACC). 

\subsection{Semi-Supervised Models}
In addition to unsupervised segmentation, TricycleGAN can be used for semi-supervsed segmentation. We implement semi-supervised learning by fine-tuning the unsupervised models with increasing amounts of labelled images. The results of this experiment are given in Figure \ref{fig:semisup}. 

\subsection{U-Net Comparison} 
We use the commonly used and widely accessible U-net \cite{Ronne2015} to train a supervised segmentation model for the ultrasound dataset. As with our U-net embedded in TricycleGAN, we train the U-net using the sum of the pixel-wise binary cross-entropy and the dice coefficient as the loss function. We use Adam for optimization with a batch size of 1. Finally, we perform data augmentation with horizontal flips (50\% probability) and horizontal and vertical translations of up to 26 pixels (10\%). 

\subsection{Mask-RCNN Comparison}
The Mask-RCNN \cite{He2017} was the segmentation model that performed best during the ISIC 2018 competition \cite{Qian2018}, therefore it is used here as a direct comparison. We use anchor sizes of $2^{i}, i\in\{3, 4, 6, 7, 8\}$ and 32 training ROIs per image, and other hyperparameters were kept to recommended values. We perform data augmentation with both horizontal and vertical flips (50\% probability), rotation of $\pi/2$ or $3\pi/2$, and a Gaussian blur of up to 5 standard deviations. Note that is not the same as the winning ISIC 2018 model that is desribed in Section \ref{sec:isic_winner}, which includes an additional encoder-decoder model for segmentation. 

\subsection{W-net Comparison}
To compare TricycleGAN with a popular unsupervised approach, we train W-net with the soft normalized cut term in the loss function \cite{Xia2017}. In addition, we perform the recommended post-processing of the W-net generated segmentation maps using a fully-connected CRF for edge recovery, and hierarchical image segmentation for contour grouping \cite{Arbalaez2011}.

\subsection{Comparison with ISIC Competition Winner}
\label{sec:isic_winner}
The winning method for the ISIC 2018 segmentation task competition \cite{Qian2018} uses Mask-RCNN for supervised detection of the ROI. The ROI bounding box output by Mask-RCNN is then flipped and rotated by different angles to create four copies of the ROI. Each of these is then segmented with a custom encoder-decoder network, and the average segmentation map is taken as the final predicted mask. 

\subsection{SegCM Comparison}
We test a recent method for unsupervised medical image segmentation based on the local centre of mass on our kidney dataset \cite{Aganj2018}. We perform nested 5-fold cross-validation to tune the alpha and power hyperparameters (alpha $\in \{250,500,750,1000\}$, power $\in \{1,2,3 \}$) with 240 training samples, 60 validation samples, and 138 test samples. Since this method provides a segmentation label for every discovered region of an image, we measure performance by taking the label that maximizes the Dice score on a per image basis, and accumulate these scores across all images on the test set.

\section{Results}

%
%


\subsection{Kidney Segmentation Performance}
\begin{figure}
	\centering
	\includegraphics[width=0.45\textwidth]{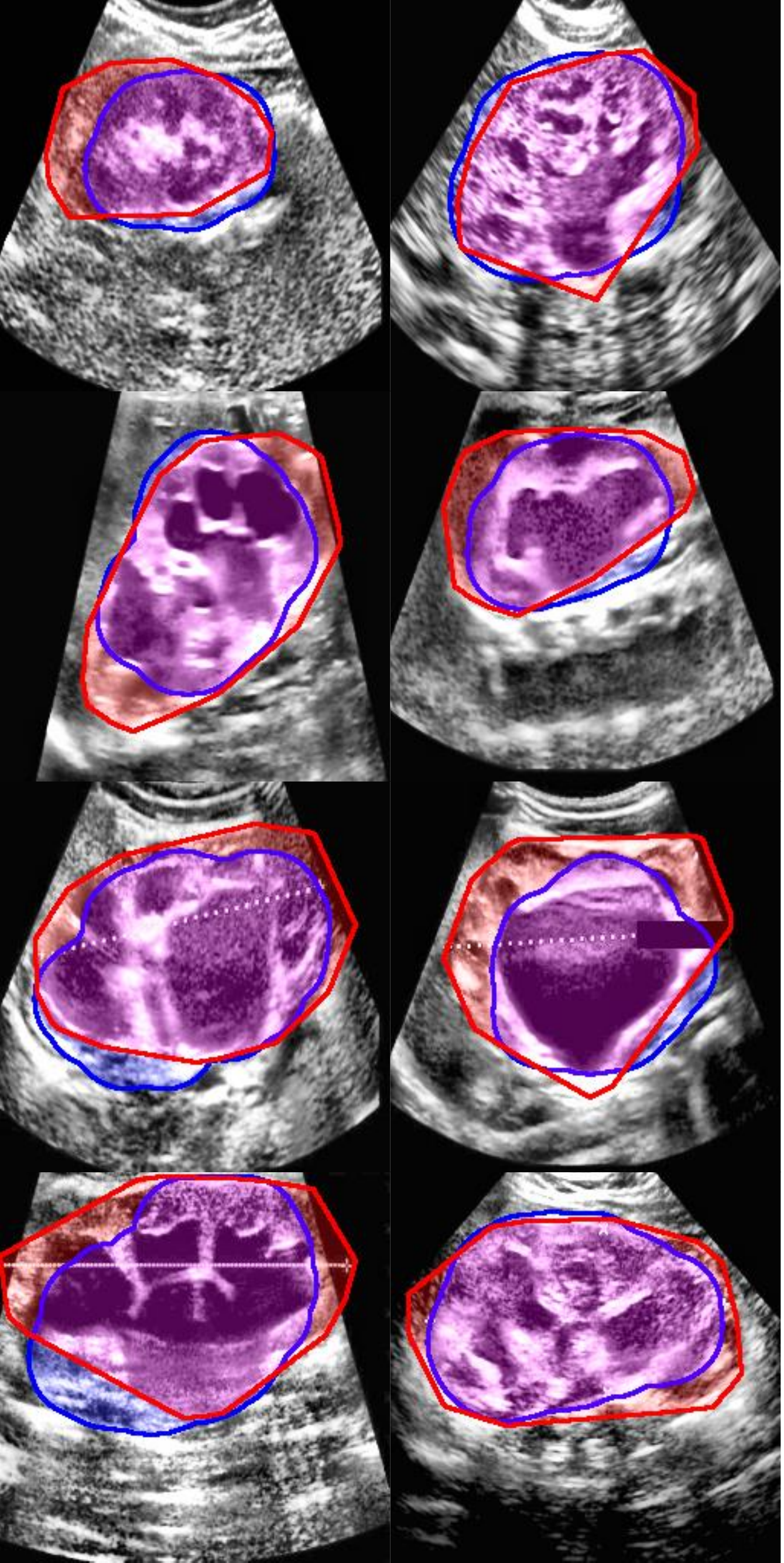}
	\caption{Kidney segmentation masks comparingTrycicleGAN (blue) to the clinician-provided ground truth labels (red) for a random subset of test images.}
	\label{fig:qual_results}
\end{figure}

In Figure \ref{fig:qual_results} we show the kidney segmentation masks produced by TricycleGAN compared with the clinician-provided ground truth for randomly selected images in the test set. In Table \ref{tab:results} we show the corresponding segmentation performance metrics. TricycleGAN trained in a semi-supervised manner using synthetic images and 45 real images and results obtained by first training the model on synthetic data and then fine-tuning it on 45 real images are listed under semi-supervised results TricycleGAN and TricycleGAN+10  respectively (with ``+10'' denoting 10\% of the labelled data used for fine-tuning; the amount required before performance plateaued). 

\begin{table}
\begin{center}
	\caption{Performance metrics for ultrasound kidney segmentation.}
	\label{tab:results}
	\begin{tabular}{| c | c | c | c | c | c | c |}
	\hline
	{}  & Model & F1 & Specificity & Sensitivity & IoU & pACC  \\ \hline
	\multirow{3}{*}{Unsup.}   
	& TricycleGAN  & 0.81 (0.10) & 0.93 (0.08) & 0.84 (0.14) & 0.69 (0.13) & 0.90 (0.06) \\ 
	& SegCM & 0.48 & 0.31 & 0.91 & 0.31 & 0.47 \\ 
	& W-net & 0.46 (0.10) & 0.20 (0.05) & 0.98 (0.02) & 0.30 (0.12) & 0.41 (0.07) \\ 
	 \hline
	\multirow{2}{*}{Semi-Sup.}
	& TricycleGAN  & 0.87 (0.11) & 0.97 (0.04) & 0.86 (0.13) & 0.78 (0.13) & 0.93 (0.05) \\ 
	& TricycleGAN+10 & 0.88 (0.08) & 0.97 (0.03) & 0.88 (0.09) & 0.80 (0.11) & 0.94 (0.04) \\ 
	\hline
	\multirow{1}{*}{Sup.}     
	& U-net & 0.91 (0.09) & 0.97 (0.04) & 0.90 (0.10) & 0.84 (0.10) & 0.95 (0.03) \\ 
	\hline
	\end{tabular}
\end{center}
\end{table}

\subsection{Skin Lesion Segmentation Performance}
\begin{figure}
	\centering
	\includegraphics[width=0.45\textwidth]{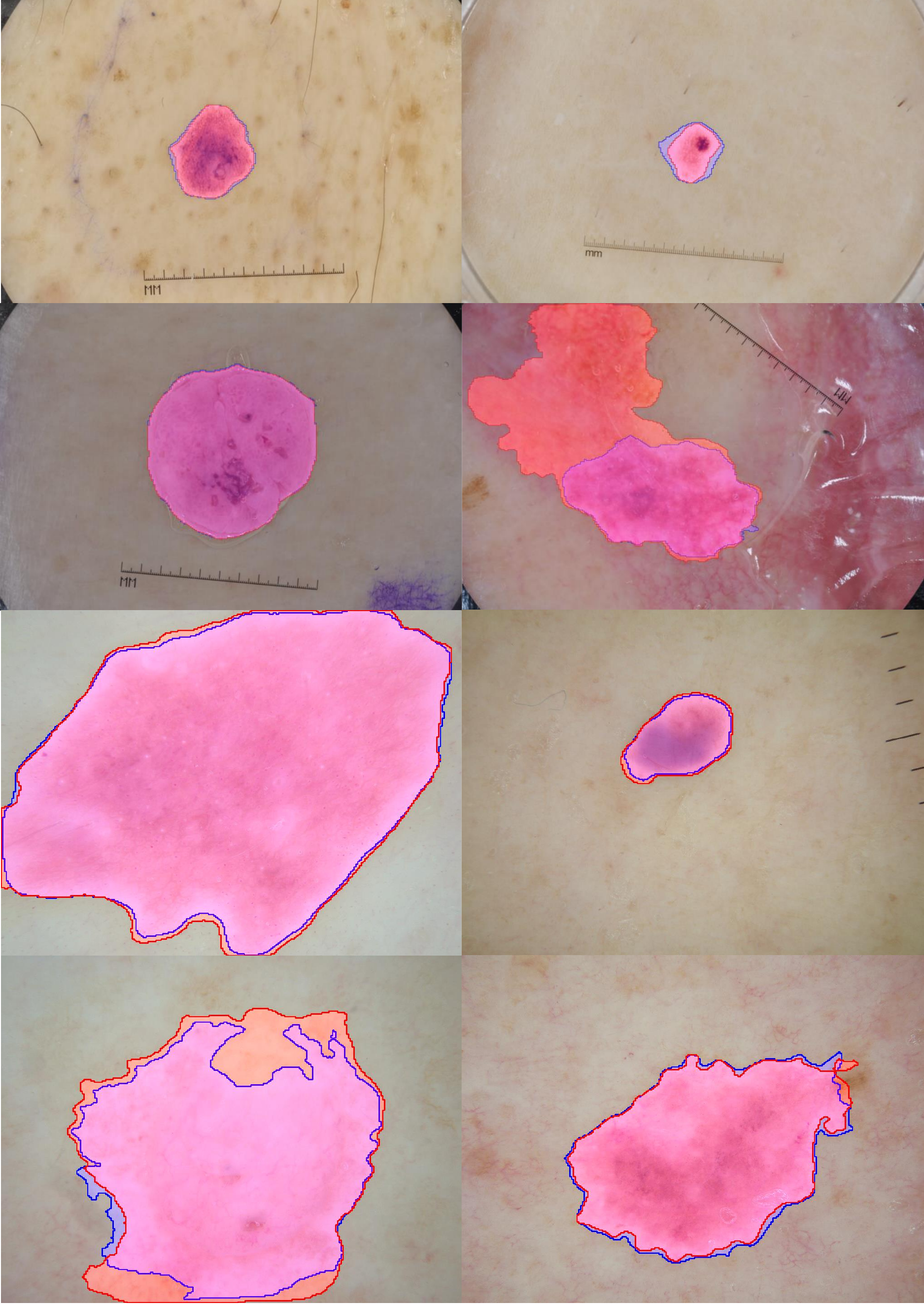}
	\caption{Skin lesion segmentation masks comparing unsupervised TricycleGAN (blue) to a semi-supervised TricycleGAN+15 (red). Images are a random subset taken from the ISIC 2018 test set, for which ground truth labels are not available for download.}
	\label{fig:qual_results_skin}
\end{figure}

Performance metrics on the ISIC 2018 dataset using TricycleGAN are shown in Table \ref{tab:results_skin} along with results obtained by the competition winner \cite{Qian2018} and current top submission. Here we use the metrics given by the online submission system, which includes a thresholded IoU (th-IoU). This metric sets all per-image IoU scores that are less than 0.65 to 0 before computing the mean IoU. Examples of the output masks on randomly selected test images are shown in Figure \ref{fig:qual_results_skin}. As with the kidney dataset, TricycleGAN+15 denotes model performance after supervised fine-tuning of the unsupervised model with 15\% of the labelled validation data. However, the submission page for model evaluation on the test set no longer provides metrics other than the th-IoU, so other metrics are not presented for this model.

\begin{table}
\begin{center}
	\caption{Performance metrics for ISIC 2018 skin lesion boundary segmentation.}
	\label{tab:results_skin}
	\begin{tabular}{| c | c | c | c | c | c | c | c |}
	\hline
	{}  & Model & th-IoU  \\ \hline
	\multirow{1}{*}{Unsup.}   
	& TricycleGAN & 0.691 \\ 
	\hline
	\multirow{1}{*}{Semi-Sup.}\ 
	& TricycleGAN+15 & 0.759 \\ 
	\hline
	\multirow{3}{*}{Sup.}     
	& Mask-RCNN & 0.763 \\ 
	& Winner \cite{Qian2018} & 0.802 \\ 
	& Current Top & 0.836 \\ 
	\hline
	\end{tabular}
\end{center}
\end{table}

\subsection{Supervised Fine-Tuning}
\begin{figure}
	\centering
	\begin{tabular}{c}
    		\subfloat[Kidney Data]{\includegraphics[width=0.35\textwidth]{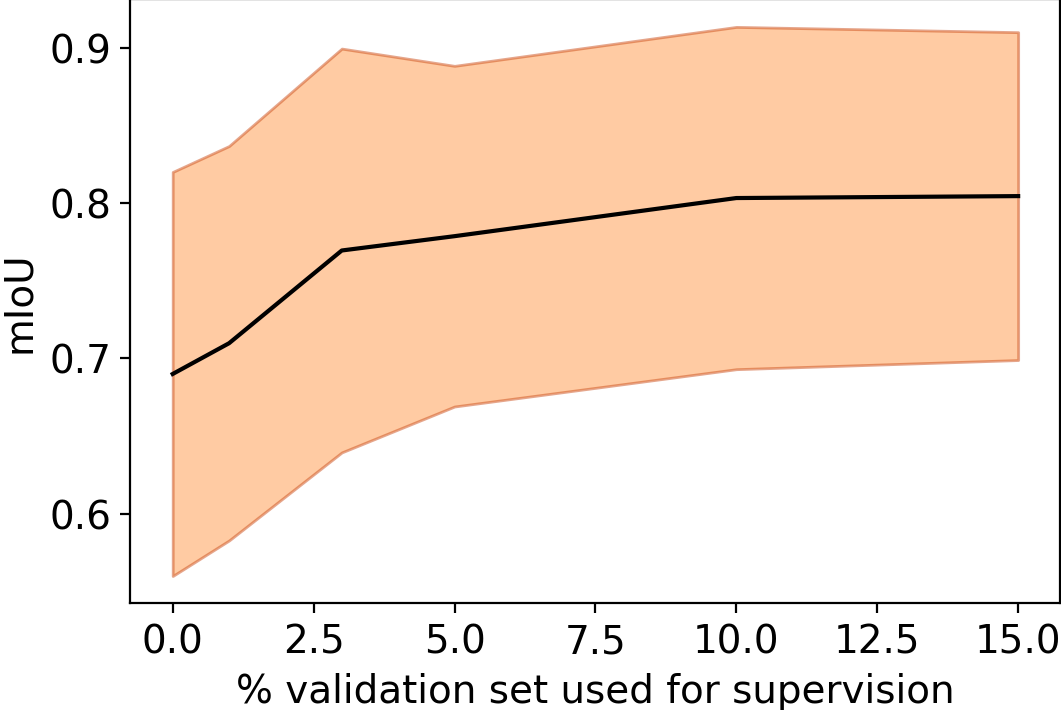}}
    		\subfloat[ISIC 2018 Data]{\includegraphics[width=0.35\textwidth]{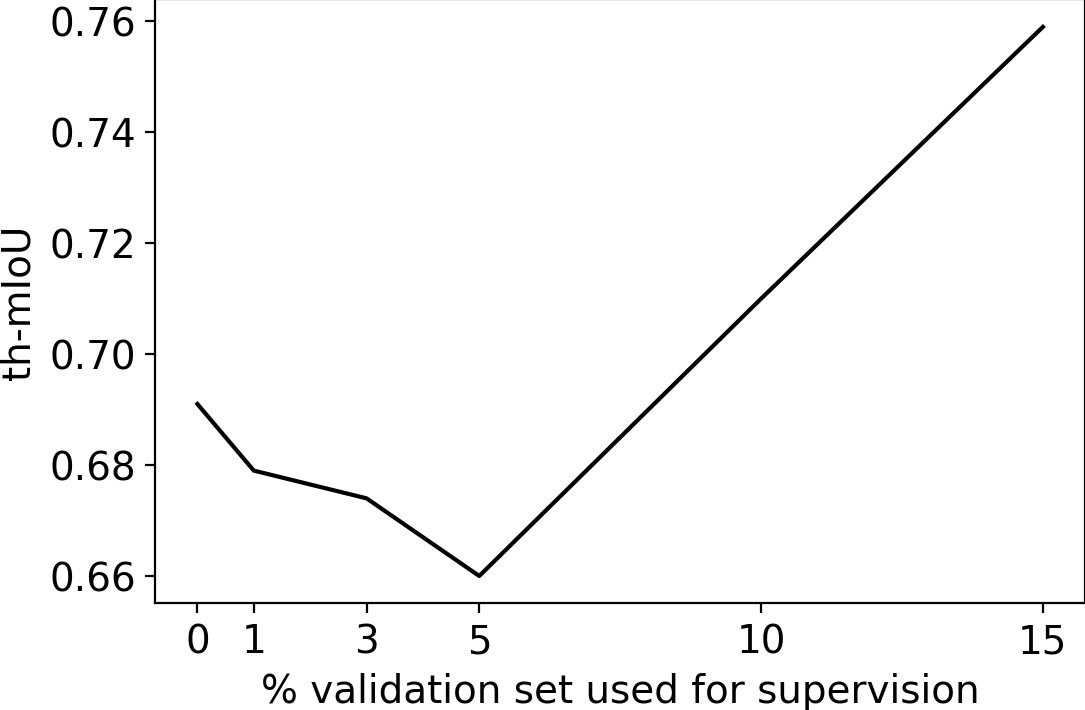}}
	\end{tabular}
	\caption{Result of fine-tuning the unsupervised models with increasing amounts of the labelled validation data.}
	\label{fig:semisup}
\end{figure}

We evaluate the effect of fine-tuning TricycleGAN with increasing amounts of labelled data. The results presented in Figure \ref{fig:semisup} suggest that only 5-10\% of the validation images are needed for optimal performance on the kidney ultrasound dataset, corresponding to 22-43 images. However, segmentation performance continues to improve on the ISIC 2018 dataset with additional data.

\section{Discussion}
We present TricycleGAN as way of creating labelled synthetic medical images from unlabelled data to enable unsupervised image segmentation. TricycleGAN takes advantage of recent advances in image synthesis and generative modelling by making use of shape priors that can be used to construct templates of an object of interest. This is accomplished by extending the recently developed CycleGAN architecture to a network comprised of three generators that translate between images, their edge maps, and their segmentation maps. 

TricycleGAN performs better than alternative unsupervised methods that do not require colour and texture features. For example, W-net performs poorly on the kidney segmentation task because it only identifies the ultrasound cone itself, rather than the kidney. We also show that our approach performs nearly as well as supervised methods for most images, though overall performance is diminished. Importantly, we show that with just a few training examples for supervised fine-tuning (here, only \~10\% of the data used for the supervised models), TricycleGAN performs close to the level of supervised models.

\bibliographystyle{ieeetran}
\bibliography{References}

\beginsupplement
\newpage
\section*{Supplementary Data}

\begin{figure}[h]
\begin{center}
\setlength{\tabcolsep}{1.5pt}
	\begin{tabular}{cc}
    		\subfloat[Grade 1]{\includegraphics[width=0.22\textwidth]{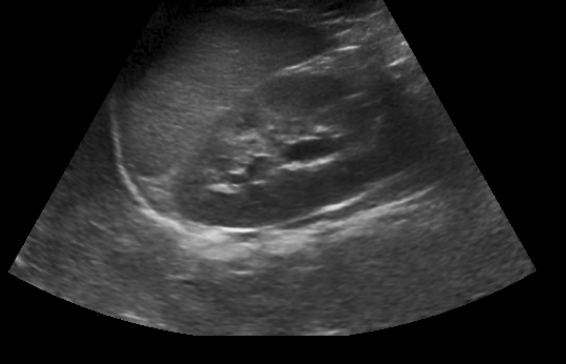}} & 
    		\subfloat[Grade 2]{\includegraphics[width=0.22\textwidth]{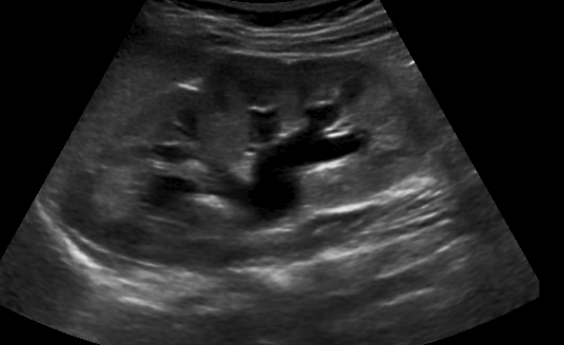}} \\[-2ex]
    		\subfloat[Grade 3]{\includegraphics[width=0.22\textwidth]{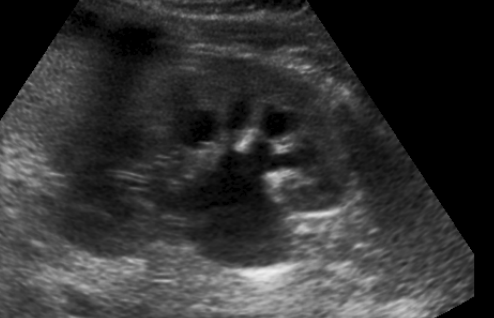}} & 
    		\subfloat[Grade 4]{\includegraphics[width=0.22\textwidth]{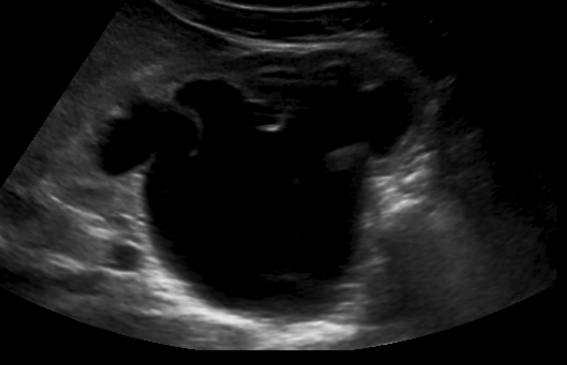}} 
	\end{tabular}
    \caption{Examples from the kidney ultrasound dataset with different hydronephrosis severity grades, from 1 (low severity) to 4 (severe hydronephrosis). }
    \label{fig:kidney_grades}
\end{center}
\end{figure}

\end{document}